\documentclass[11pt,a4paper]{article}
\usepackage{graphicx,makeidx,setspace}
\usepackage{sendms}
\usepackage{endfloat}
\begin{document}
\doublespacing
\title{Noise-enhanced computation in a model of a cortical column\\
\small{Running title: Noise-aided computation in a cortical model}
\footnote{Character count: 15250 for text + 3600 for figures}}
\author{Julien \textbf{Mayor} and Wulfram \textbf{Gerstner}\\
\\
Brain-Mind Institute and \\
School of Computer and Communication Sciences\\
Ecole Polytechnique F\'{e}d\'{e}rale de Lausanne (EPFL)\\
CH-1015 Lausanne, Switzerland}
\date{May 4, 2005}

\maketitle
\newpage
\begin{abstract}
Varied sensory systems use noise in order to enhance detection of
weak signals. It has been conjectured in the literature that this
effect, known as \emph{stochastic resonance}, may take place in
central cognitive processes such as the memory retrieval of
arithmetical multiplication. We show in a simplified model of
cortical tissue, that complex arithmetical calculations can be 
carried out and are enhanced in the presence of a stochastic background. 
The performance is shown to be positively correlated
to the susceptibility of the network, defined as its sensitivity
to a variation of the mean of its inputs. For nontrivial arithmetic tasks such 
as multiplication, stochastic resonance is
an emergent property of the microcircuitry of the model network.

\\
\emph{Contact author: Wulfram.Gerstner@epfl.ch}

\end{abstract}
\newpage

\section{Introduction}
Noise is usually considered as having a corrupting effect on
meaningful signals. There is however one well known counter
example to this widespread belief; the \emph{stochastic
resonance} (SR) phenomenon. In this case, addition of a random
interference signal (noise) to a weak, subthreshold stimuli, may
enhance its detection. Originally introduced in the framework of
physics
\cite{Benzi81,Wiesenfeld98,Collins96,Gammaitoni95,Gammaitoni98,McNamara89,Wiesenfeld94,Wiesenfeld95},
stochastic resonance is now known to take place in several sensory
systems from the cricket cercal sensory system \cite{Levin96}, to
crayfish mechanoreceptors \cite{Douglass93}, the somatosensory
cortex of the cat \cite{Mori02} and the human visual cortex
\cite{Kitajo03,Hidaka00} in order to facilitate detection of weak
signals (for a review see \cite{Moss04}).

In addition there is evidence that stochastic resonance may be
used not only at the level of sensory processing, but also in
central cognitive processes. In a psychophysical study by Usher
and Feingold, the memory retrieval of arithmetical multiplication
was found to be enhanced by the addition of noise \cite{Usher00}.

In this article, we feed a simplified model of a cortical column
with two time-varying input signals. We first show that for a
broad set of computation based on those input signals, addition
of a small amount of noise enhances the computational power of the
system. The location where the noise strength is optimal lies
approximately where the network reacts the strongest to a change
in the mean input (maximum susceptibility). We then set the
connectivity to zero (with appropriate scaling of the statistics
of the input to the neurons). Although the simplest task
(addition of both input signals) can be achieved with a similar
accuracy to that obtained with a connected network, a
multiplicative task can only be solved if the population of
neurons is connected. The stochastic resonance effect is thus
seen to take place at the \emph{system level} \cite{Masuda02}
rather than at the single cell level. It is an emergent property
of the neuronal assemblies.

\section{Methods}

We consider in our simulations networks of $N=200$ cells (leaky
integrate-and-fire neurons). The connectivity matrix is fixed and
every neuron receives input from $C_E=40$ excitatory and $C_I=10$
inhibitory presynaptic neurons randomly chosen among the $N-1$
neurons in the network. The system is made out of $80\%$
excitatory and $20\%$ inhibitory neurons, reflecting the ratio of
pyramidal cells to interneurons in cortical tissue. This excess of
excitatory neurons is approximately balanced by the greater
efficacy of synaptic transmission for inhibition; in our model
six times bigger than for excitation: $\omega_I=7.2$mV and
$\omega_E=1.2$mV. This approximate balance between excitation and
inhibition is thought to take place at a functional level in
cortical areas \cite{Liu04,Shu03}. Sparsely-connected networks of
spiking neurons of this type have been fully described in terms
of their dynamical behaviour \cite{Brunel00}. The dynamics of the
leaky integrate-and-fire neurons is described by the following
equation:
\begin{equation}\label{if}
    \tau_m \dot{u_i}(t) = -u_i(t) + RI_i^{netw}(t) + RI_i^{ext}(t)
\end{equation}
where $u_i(t)$ describes the membrane potential of neuron $i$
with respect to its equilibrium value $u_i=0$, $\tau_m=20$ms
corresponds to its effective membrane time constant and
$R=100M\Omega$ is the effective input resistance of the neuron
stimulated by a total input $RI_i^{netw}(t) + RI_i^{ext}(t)$. We
add to this equation a threshold condition; if the membrane
potential of the neuron exceeds the critical value $\theta=20$mV,
a spike is emitted and after a refractory period $\tau_{rp}=2$ms,
the integration start again from the reset potential
$u_{reset}=0$mV. Every time the neuron $i$ receives an action
potential from a presynaptic pyramidal cell $j$ (resp.
interneuron), its membrane potential is depolarized according to
the value of the synaptic efficacy for excitation
$\omega_j=+\omega_E$ (resp. for inhibition, the neuron is
hyperpolarized by an amount $\omega_j=-\omega_I$). The input a
neuron $i$ will get from within the network can thus be written
as:
\begin{equation}\label{synint}
    RI_i^{netw}(t)=\tau_m \sum_{j \in M_i} \omega_j \sum_k \delta(t-t_j^k-D)
\end{equation}
where $M_i$ is the ensemble of presynaptic neurons,
$t_j^k$ the time neuron $j$ fires its $k$'th spike
and $D=1$ms is a short transmission delay.

We can now decompose the external stimulation $RI_i^{ext}(t)$
into two contributions. First, we model all noise sources,
ranging from synaptic bombardment from neurons outside the network
to different sources of noise diversely located at the level of
synaptic transmission, channel gating, ion concentrations,
membrane conductance to name but a few. All these noise sources
are grouped into a term $RI^{backgr}$ with a mean depolarisation
$RI^{dep}$ and a white noise component defined by its standard
deviation $RI^{noise}$. We can define the \emph{susceptibility}
$\chi$ of the network as the sensitivity of the population
spiking rate $\nu$ upon a change in the mean depolarisation
$RI^{dep}$: $\chi=\frac{\partial \nu}{\partial I^{dep}}$. We
constrained our simulation space to small mean depolarisation
$RI^{dep}$ so that in absence of the noisy component
$RI^{noise}(t)$, the network exhibits no spiking activity
(subthreshold regime).

\begin{figure}[!hbt]
  \centering
  \includegraphics[width=8.2cm]{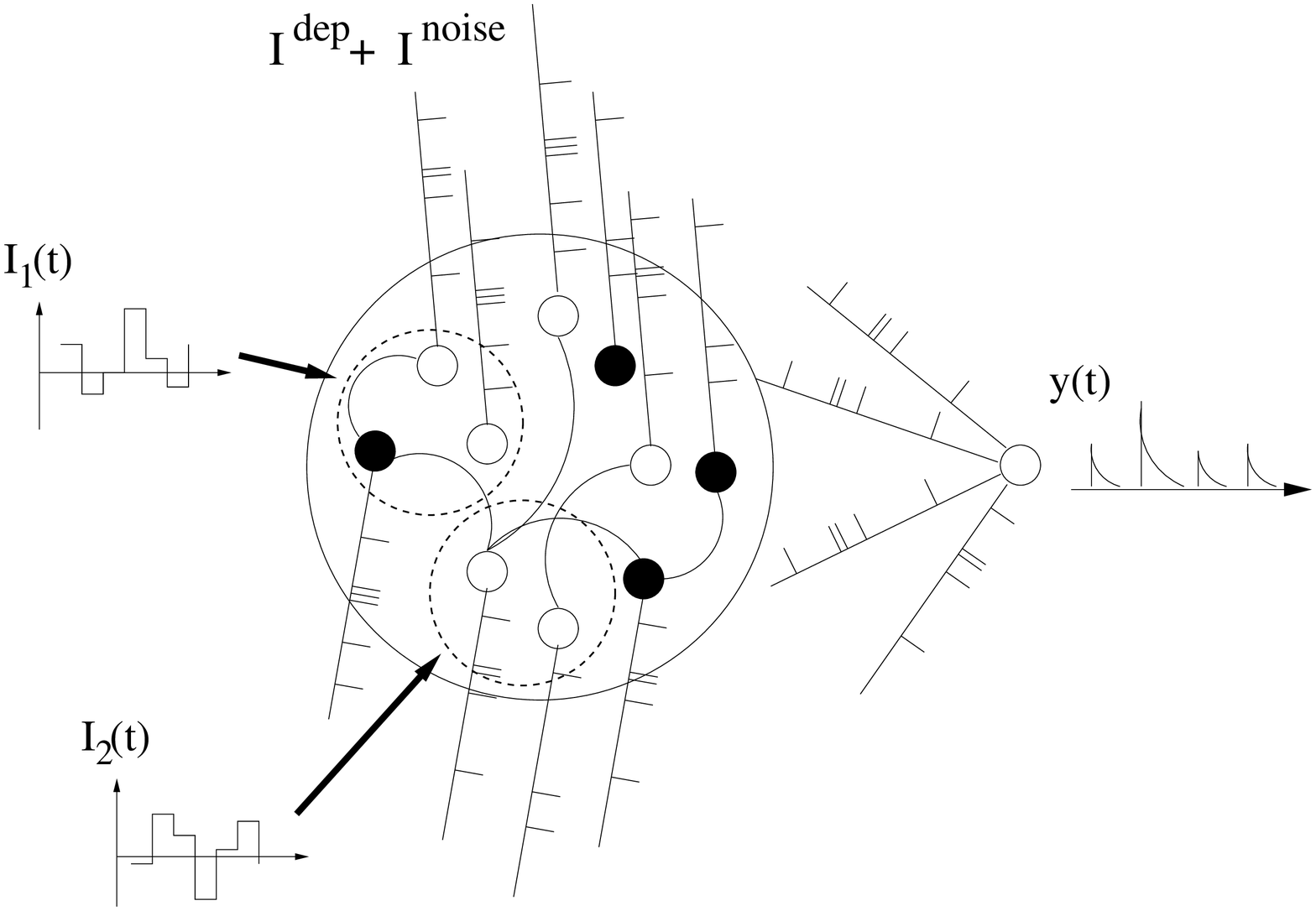}
  \caption{Structure of the network. All neurons in the network receive in addition to their recurrent afferents
  a background stochastic input made out of $I^{dep}$ and $I^{noise}(t)$. In addition two randomly generated
  subpopulations, each composed of $20\%$ of the total number of neurons, receive test inputs $I_1(t)$ and $I_2(t)$.
  The readout $y(t)$ \emph{sees} all neurons in the network and is trained to match a function of the test input $I_1$
  and $I_2$ (see the methods section for details).
  }\label{network}
\end{figure}

Then, we inject two test signals $I_1(t)$ and $I_2(t)$, each to
$20\%$ randomly chosen neurons in the network. Both inputs share
the same statistical properties; they are constant over a time
interval $T=40$ms and then they switch to new randomly chosen
values and remain constant for the next time interval $T$. At
each transition time, the new values are chosen uniformly over
the interval $[-50pA,50pA]$. We want to know how performance in a
series of computational tasks based on both test signals $I_1(t)$
and $I_2(t)$ are affected by the noise level. We adapt a paradigm
introduced in the framework of Liquid State Machines
\cite{MaassETAL:01a} or Echo State Networks \cite{Jaeger04}: we
consider a readout with a dynamics $dy/dt =
-((y-\alpha_0)/\tau_s) + \sum_{i=1}^N \alpha_i \sum_k
\delta(t-t_i^k)$ where the sums run over all firing times $t_i^k$
of all neurons in the network. $\tau_s=5$ms is a short synaptic
time constant. The $N+1$ free parameters $\alpha_i$ ($0\le i\le
N$) are chosen so as to minimize the signal reconstruction error
between the readout and the target: $E = \langle [y(t) -
F(I_1(t-\Delta),I_2(t-\Delta))]^2\rangle$. In our simulations we
fixed $\Delta=15ms$ so that the transient period after a
transition has vanished. The set of functions $F(I_1,I_2)$ that
are under consideration in the present study are; the addition
$F=I_1+I_2$, the multiplication $F=I_1I_2$ which plays a crucial 
role in the transformation of object locations 
from retinal to body-centered coordinates \cite{Andersen90} 
and two polynomials of degree two $F=(I_1+I_2)^2=I_1^2+I_2^2+I_1I_2$ 
and $F=(I_1-I_2)^2=I_1^2+I_2^2-I_1I_2$, which can be seen as the 
nonlinear XOR paradigm \cite{Minsky69}.

Parameters were optimized using a first simulation (learning set)
lasting 100 seconds (100'000 time steps of simulation) and were
kept fixed afterwards. The performance measurements reported in
this paper are then evaluated on a second simulation of 100
seconds (test set). We compare our results to a simple two
parameter readout. Such a readout only adjusts to the mean of the
time series $F(I_1(t),I_2(t))$. The reconstruction error for such
a readout equals the variance $\sigma_{\rm F}^2$ of the time
series $F(I_1(t),I_2(t))$. The performance with the full readout
are therefore expressed as a gain (in percent) over the trivial
prediction: $G=100(1-\frac{E}{\sigma_{\rm F}^2})$, where $E$ is 
the error introduced above.

In the simulation of Fig.\ref{control}, all connections in the
network were removed. In order to compensate the loss of input
$I^{netw}$, we changed the background input $RI^{backgr}$ so that
the neurons receive an input with the same statistical properties
(mean and variance). The adapted input in the network with 
''no connection'' (nc) is thus:
$RI^{dep}_{nc}=RI^{dep}+\nu \tau_m(C_E \omega_E - C_I \omega_I)$
for the mean and $RI^{noise}_{nc}=RI^{noise}+\sqrt{\nu \tau_m
(C_E \omega_E^2 + C_I \omega_I^2)}$ for the variance term. $\nu$
is the mean population rate in the connected network (see
\cite{Brunel00}).

Simulation results were obtained using the simulation software
NEST\footnote{NEST Initiative, available at www.nest-initiative.org}.

\section{Results}

\begin{figure}[!hbt]
  \centering
  \includegraphics[width=11cm]{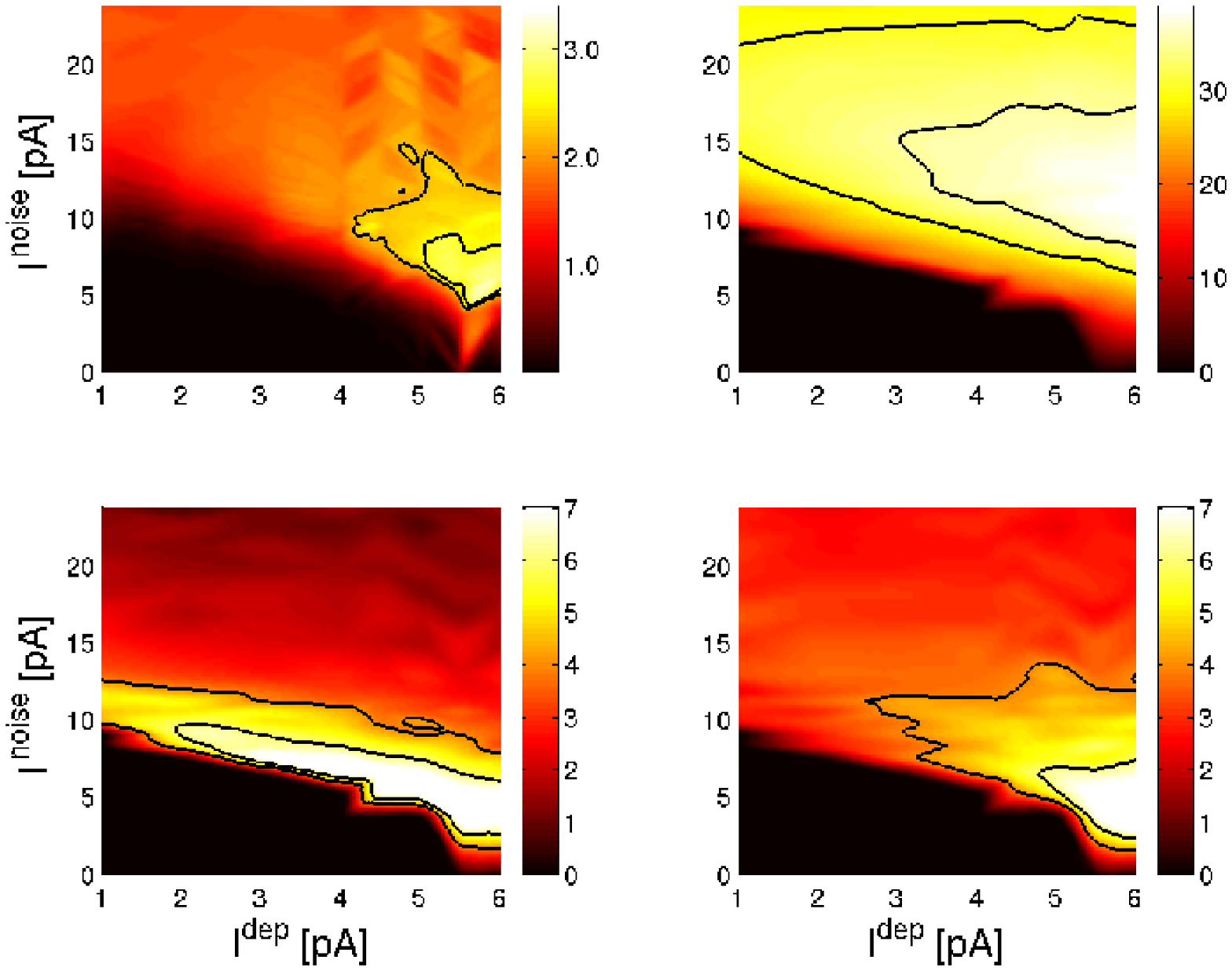}
  \caption{Top left: \emph{susceptibility} $\chi=\frac{\partial \nu}{\partial I^{dep}}$ 
  in (Hz/pA) of the network as a function of the statistical properties
  of the external drive (mean and variance). A high
  \emph{susceptibility} (light colors) defines a network that is highly sensitive
  to a change in the mean of the drive.
  Top right and bottom: gain (in percent) over the trivial prediction as a function of the mean and variance of the external drive;
  for the simple additive task $I_1+I_2$ (top right), for a first
  polynomial of degree two $I_1^2+I_2^2+2I_1I_2$ (bottom left) and a second polynomial of degree two $I_1^2+I_2^2-2I_1I_2$
  (bottom right). Level curves are shown for the sake of clarity at $4\%$ and $6\%$ for the polynomials and
  at $30\%$ and $35\%$ for the addition.
  The system exhibits stochastic resonance for all three different
  tasks. In addition, the location where the performance peak approximately
  corresponds to the zone of high \emph{susceptibility}.}\label{gradient}
\end{figure}

We want to know how noise influences the ability of a recurrent
network of spiking neurons to process information and to perform
a series of computational tasks. In particular we are interested 
in effects related to stochastic resonance. We therefore 
relate the optimal noise level (where the system has a maximum
performance) to dynamical properties of the network. From that
perspective, we analyze how performance in a series of
computational tasks based on both test signals $I_1(t)$ and
$I_2(t)$ are affected by the noise level. In a first series of
simulations, we measure the gain over the trivial prediction for
three different functions of the test signals; $F=I_1+I_2$,
$F=(I_1+I_2)^2$ and $F=(I_1-I_2)^2$ (see figure \ref{gradient},
respectively top right, bottom left and bottom right graphs).
Note that the function $F=(I_1-I_2)^2$ can be seen as an
implementation of the XOR task, which cannot be solved by a
single layer neural network (perceptron) \cite{Minsky69}. In all
three tasks, the network exhibits stochastic resonance; for a
given mean depolarisation $RI^{dep}$, there is a non-monotonic
dependence upon the noise level. The maximum gain compared to 
trivial prediction reaches $38\%$ for the additive task and
$7-9\%$ for polynomials of degree two. A comparison to the map of
the susceptibility of the network $\chi=\frac{\partial
\nu}{\partial I^{dep}}$ (see top left graph of figure
\ref{gradient}) indicated that the tasks are solved best when the
sensitivity of the network upon changes in the mean input is
highest. Addition of noise both increases the susceptibility of
the network and the capacity of performing complex computation
based on sparse inputs.

In order to know whether stochastic resonance displayed in the
network is an emergent property of the population of neuron
\cite{Masuda02} or a single cell effect, we removed all
connections within the network. In this second series of
simulations, we compare the performance achieved in networks with
connectivity to networks with no connectivity. The latter
networks receive adapted version of the mean input
$RI^{dep}_{nc}$ and of the noisy input $RI^{noise}_{nc}$ so that
every neurons is stimulated with a mean and variance equivalent
to that of the connected network (see methods).

\begin{figure}[!hbt]
  \centering
  \includegraphics[width=8.2cm]{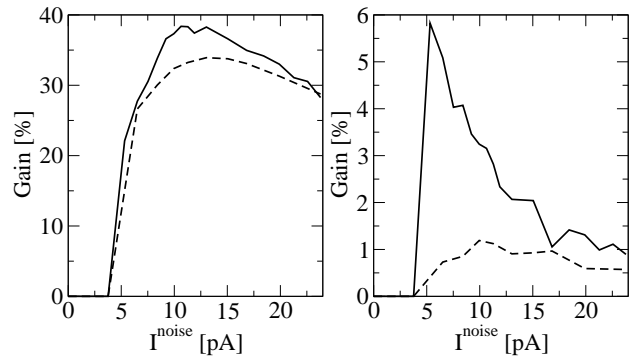}
  \caption{Left: Gain (in percent) over the trivial prediction
  for the additive task $I_1+I_2$; for the connected network
  (solid) and for the control network, when the connectivity is set to
  zero (dashed). $I^{dep}=5.5$pA.
  The unconnected collection of neurons is capable of solving this simple task with a similar
  accuracy as the randomly connected neural network.
  Right: Gain (in percent) over the trivial prediction for the multiplicative task
  $I_1I_2$; for the reference network (solid) and when the connectivity is set to zero (dashed).
  In absence of recurrence, the network is no longer able to sustain complex computations.
  Stochastic resonance is thus a \emph{population-based} effect rather than a \emph{single-cell} phenomenon.}\label{control}
\end{figure}

In the first task; the addition $F=I_1+I_2$, the performance with
and without connectivity are similar (see figure \ref{control}
left). Since the readout unit performs a weighted \emph{sum} that
runs over all neurons, it can capture the essence of this simple
computation by summing the averaged response of the groups of
neurons receiving input $I_1$ with the averaged response of those
receiving $I_2$. Recurrent loops play here no significative role.
In the second task we train the network to perform the
multiplication of the two test signals $F=I_1I_2$. This
arithmetical operation is thought to be essential to the brain in
order to do coordinate transformation \cite{Andersen90}. The computation of a
multiplication by a recurrent neural network was shown to be
achievable in a model of the parietal cortex \cite{Salinas96, Pouget97}. In
this multiplicative task $F=I_1I_2$, the complex recurrent
network outperforms the network with no connectivity (see figure
\ref{control} right). In fact, in absence of connections within
the population of neurons, multiplication cannot be solved by the
simple addition of noise. Stochastic resonance displayed in the
multiplicative task therefore takes place at a \emph{system}
level rather than at the level of single neurons.

\section{Conclusion}

Complex networks of neurons fall in the class of non
linear systems with a threshold; systems that are known to
exhibit stochastic resonance. 
From the experimental side, evidences have shown
that the phenomenon helps in detecting sensory signals of small
amplitude, and furthermore to favor high level cognitive
processes such as arithmetical calculations. 
Our model has revealed the presence of stochastic resonance in a series 
of neural-based computation. 
Whereas simple
additive transformations of input signals can be solved 
by a collection of independent neurons, more complex
computations need the massive
recurrence typically observed in cortical tissue.
Such complex tasks include the XOR problem, a nonlinear benchmark test,
and the arithmetical multiplication the brain is likely to use in order
to achieve coordinate transformation. Stochastic
resonance displayed is then an emergent property of the brain
microcircuitry.

\end{document}